\documentclass[a4paper,11pt]{article}
\usepackage{jheppub}
\usepackage[dvipdfmx]{color}
\usepackage[dvipdfmx]{graphicx}
\usepackage{amsfonts}
\usepackage{amssymb}
\usepackage{amsmath}
\usepackage{bm}

\def\gsim {~^{>~}_{\sim~}}

\usepackage{ulem}
\usepackage{color}

\title{Two dimensional gluon propagators
in maximally Abelian gauge in SU(2) Lattice QCD}

\author[a,b]{Shinya~Gongyo}
  \affiliation[a]{Department of Physics, New York University,
   4 Washington Place, New York, New York 10003, USA}
  \affiliation[b]{Department of Physics, Graduate School of Science,
  Kyoto University, Kitashirakawa-oiwake, Sakyo, Kyoto 606-8502, Japan}

\emailAdd{gongyo@ruby.scphys.kyoto-u.ac.jp}

\date{\today}
\abstract{
Using SU(2) lattice QCD in two dimensions, we study diagonal and off-diagonal gluon propagators in the maximally Abelian gauge (MAG) with U(1)$_3$ Landau gauge fixing. These propagators are investigated both in momentum space and coordinate space. The Monte Carlo simulation is performed at $\beta=7.99, 30.5,$ and $120$ on $62^2, 128^2,$ and $256^2$ at the quenched level.
In the momentum space, the transverse component of the diagonal gluon propagator shows suppression with increasing $\beta$ in the infrared region and the dressing function at $\beta=120$ has a maximum at $p^2 \simeq 4$GeV, while the transverse component of the off-diagonal gluon propagator does not show the $\beta$-dependence and the dressing function does not have a maximum. This behavior indicates that the effect of the Gribov copies is found for the diagonal gluon, consistent with the result obtained by the Gribov-Zwanziger action in the MAG. In addition, this behavior supports that the Abelian dominance is not found in two dimensions. 
 In the coordinate space, the diagonal gluon propagator decreases as $\beta$ increases at long distance. In particular, at $\beta=120$ the diagonal propagator decreases more rapidly with increasing distance than the off-diagonal propagator. These behaviors also indicate the presence of Gribov copies and no Abelian dominance in two dimensions.
Furthermore, we also study these propagators at zero-spatial-momentum. The result suggests that all of the spectral functions of diagonal and off-diagonal gluons would have negative regions and thus they show the violation of the Kallen-Lehmann representation.
}

\begin{document}
\maketitle
\section{Introduction}

The behaviors of the gluon propagator and the ghost propagator in the infrared region play an important role for understanding color confinement in quantum chromodynamics (QCD). According to scenarios devised by Kugo and Ojima, and Gribov and Zwanziger, confinement is related to the (deep) infrared behavior of the gluon propagator and the ghost propagator in the Landau gauge \cite{Kugo:1979gm,Gribov:1977wm,Zwanziger:1989mf,Zwanziger:1991gz}.
Gribov and Zwanziger suggested that, because of the effects of the Gribov copies, the functional form of the gluon propagator is largely changed and shows the violation of the Kallen-Lehmann representation.  
From this point of view, the gluon propagator in the Landau gauge has been studied using lattice QCD calculations and analytical frameworks \cite{Alkofer:2000wg,Fischer:2006ub,Vandersickel:2012tz,Pawlowski:2005xe}.

Recently, the accurate lattice studies have reported that the infrared behavior of the gluon propagator in the Landau gauge greatly depends on the dimensionality \cite{Vandersickel:2012tz,Cucchieri:2007md,Cucchieri:2007rg,
Bogolubsky:2009dc,Maas:2007uv,Maas:2011se,Dudal:2010tf,Oliveira:2012eh}. In both four and three dimensions, the gluon propagator would not vanish at zero momentum \cite{Cucchieri:2007md,Bogolubsky:2009dc,Maas:2011se}, which does not seem to be consistent with the Gribov-Zwanziger scenario and Kugo-Ojima scenario. In order to solve the challenge, the refined Gribov-Zwanziger action, which introduces condensates into the original Gribov-Zwanziger action, has been constructed \cite{Dudal:2008rm,Dudal:2008sp,Dudal:2007cw,Vandersickel:2012tz}. On the other hand, in two dimensions, the lattice studies show that the gluon propagator seems to vanish at zero momentum \cite{Cucchieri:2007rg, Maas:2007uv}. This indicates that the original Gribov-Zwanziger action is working well in two dimensions, and the scaling solution in Schwinger-Dyson equations, which goes to zero as momentum goes to zero in two dimensions, is also in agreement with the lattice result \cite{Zwanziger:2001kw}. In addition to these studies, the fact in two dimensions that the gluon propagator goes to zero with decreasing momentum is shown analytically \cite{Zwanziger:2012xg}, and the proof is also supported by a numerical study \cite{Gongyo:2014jfa}.

Also in the maximally Abelian gauge (MAG), the gluon propagator in four dimensions has been investigated using lattice calculations \cite{Gongyo:2012jb, Gongyo:2013sha, Bornyakov:2003ee,Amemiya:1998jz} and analytical frameworks \cite{Schaden:1999ew,Huber:2009wh, Mader:2013ru, Kondo:2011ab, Dudal:2004rx} from the viewpoint of the dual-superconductor picture proposed by Nambu, 't Hooft and Mandelstam \cite{Nambu:1974zg,Mandelstam:1974pi}. The lattice studies in four dimensions show that the diagonal gluon propagator is more largely enhanced than that of the off-diagonal propagator at low momentum, which means that the diagonal gluon plays an important role for the infrared physics (so-called Abelian dominance) \cite{'tHooft:1981ht,Ezawa:1982bf,Ezawa:1982ey}. The scaling solution in Schwinger-Dyson equations in the MAG also supports the behavior \cite{Huber:2009wh}. Furthermore, many studies support that QCD vacuum in the four dimensional MAG is in good agreement with the dual superconductor in the MAG \cite{Kronfeld:1987vd, Kronfeld:1987ri, Suzuki:1989gp, Brandstater:1991sn, Stack:1994wm, Miyamura:1995xn, Woloshyn:1994rv}. 

Whether the Abelian dominance is supported or not by the gluon propagator in two dimensions is of interest. According to the dual-superconductor picture, the monopole condensation leads to confinement. In four-dimensional MAG, the topology is given by $\Pi _2$ (SU(2)/U(1)$_3)\simeq $ {\bf Z} and monopoles may appear. However, in lower dimensions, no monopole seems to appear. Therefore, there is no reason to support the Abelian dominance in two dimensions.

In addtion, the behavior of the two dimensional gluon propagator in MAG is of significance from the perspective of the Gribov copies \cite{Bruckmann:2000xd,Bali:1996dm} as in the case of the Landau gauge. Recently, the Gribov-Zwanziger action in the Landau gauge has been generalized to other gauges such as the maximally Abelian gauge (MAG) \cite{Capri:2005tj,Capri:2006cz,Capri:2008ak,Capri:2008vk,Capri:2010an,Gongyo:2013rua} and the R$_\xi$ gauge \cite{Lavrov:2011wb,Lavrov:2013boa,Moshin:2014xka}. In the MAG, the presence of the Gribov region leads to an anomalous behavior of the diagonal propagator at tree level which means the violation of the Kallen-Lehmann representation, while the off-diagonal propagator is not affected.

The aim of this paper is to investigate the gluon propagator in two dimensions in the MAG with the U(1)$_3$ Landau gauge in momentum space and coordinate space.
In Sec. \ref{2}, we briefly summarize the definition of the MAG with the U(1)$_3$ Landau gauge, and the diagonal and off-diagonal gluon propagators. We investigate the momentum-space gluon propagators in Sec. \ref{4} and the coordinate-space gluon propagators in Sec. \ref{3}. We show the summary in Sec. \ref{5}.

\section{SU(2) maximally Abelian gauge  with U(1)$_3$ Landau gauge and the gluon propagators}
\label{2}
The MAG on SU(2) lattice is given by the maximization of
\begin{align}
R_{\rm MA}&\equiv \sum_x \sum^d_{\mu =1} {\rm tr} \left[ U_\mu (x) \sigma ^3 U_\mu^{\dagger}(x) \sigma^3  \right],\label{MA}
\end{align}
where $d$ is the dimension of Euclidean space-time, $\sigma^a (a=1,2,3)$ is the Pauli matrix and $U_\mu (x)$ is the SU(2) link-variable, $U_\mu (x)=\exp\left[\sum_{a=1,2,3}iA^a_\mu (x)\sigma ^a/2 \right] \in $SU(2) with the lattice gauge field $A^a$.

This condition is invariant under the U(1)$_3$ gauge transformation,
\begin{align}
U_\mu (x) \rightarrow h(x) U_\mu (x) h^\dagger (x+\mu)
\end{align}
with $h(x)= \exp[i\theta ^3 \sigma ^3 /2]$ and does not fix the gauge symmetry.
Using the Cartan decomposition, $U_\mu (x)\equiv M_\mu (x) u_\mu (x)$ with $M_\mu (x) \in $SU(2)/U(1)$_3$ and $u_\mu(x) \in$ U(1)$_3$, the residual gauge symmetry is fixed by the maximization of 
\begin{align}
R_{\rm U(1)_3L}\equiv \sum_x \sum_{\mu=1}^d {\rm tr}[u_\mu(x)]. \label{U1Landau}
\end{align}
This condition leads to the Landau gauge for the U(1)$_3$ symmetry in continuum limit, $\partial_\mu A_\mu ^3 = 0$. This procedure for the gauge fixing chooses some Gribov copy in the Gribov region in the MAG \cite{Bruckmann:2000xd,Bali:1996dm}.
 
We use the gauge fields extracted directly from the link-variables without expanding the exponential:
\begin{align}
A_\mu^a (x) &=\frac{2U_\mu ^a}{ag\left(\sum_{a=1,2,3}U_\mu ^a U_\mu ^a\right)^{1/2}}\arctan \frac{\left(\sum_{a=1,2,3} U_\mu ^a U_\mu ^a\right)^{1/2}}{U_\mu ^0} \notag \\
&=\frac{\beta^{1/2}U_\mu ^a}{\left(\sum_{a=1,2,3}U_\mu ^a U_\mu ^a\right)^{1/2}}\arctan \frac{\left(\sum_{a=1,2,3} U_\mu ^a U_\mu ^a\right)^{1/2}}{U_\mu ^0},
\end{align}
where we used $U_\mu (x) = U_\mu ^0 (x) + i \sigma ^a U_\mu ^a (x)$ and $\beta\equiv 4/(a^2g^2)$.
Using this definition, the diagonal and off-diagonal gluon propagators are given by
\begin{align}
G_{\mu\nu}^{\rm diag}(x-y) &\equiv  \left< A_\mu^3(x)A_\nu^3(y)\right>,  \notag \\
G_{\mu\nu}^{\rm off}(x-y) &\equiv \frac{1}{2} \sum_{a= 1,2} \left< A_\mu^a(x)A_\nu^a(y)\right>.				\label{def_G}
\end{align}

The momentum-space gluon propagators on the periodic lattice of $L_1 \times L_2\times \times\dots L_{d-1}\times L_{d}$ are given by
 \begin{eqnarray}
G_{\mu\nu}^{\rm diag}(p) \equiv \left< \tilde{A}_\mu^3(\tilde{p}) \tilde{A}_\nu^3(-\tilde{p})\right>,   \label{G_p^diag} \\
G_{\mu\nu}^{\rm off}(p) \equiv
\frac{1}{2} \sum_{a= 1,2} \left< \tilde{A}_\mu^a(\tilde{p})\tilde{A}_\nu^a(-\tilde{p})\right>,
				\label{G_p^off}
\end{eqnarray}
where $\tilde{p}$ and $p$ are defined as
\begin{eqnarray}
\tilde{p}_\mu\equiv \frac{2\pi n_\mu}{aL_\mu},~ p_\mu \equiv \frac{2}{a}\sin \left( \frac{\tilde{p}_\mu a}{2} \right) ,
\end{eqnarray}
with a the lattice spacing and $n_\mu = 0,1,2, \ldots ,L_\mu -1$, 
and $\tilde{A}^a_\mu (\tilde{p})$ is defined as
 \begin{eqnarray}
\tilde{A}^a_\mu (\tilde{p})  = \frac{1}{V^{1/2}}\sum _x e^{-i\sum_\nu \tilde{p}_\nu x_\nu -\frac{i}{2}\tilde{p}_\mu} A^a_\mu (x)
\end{eqnarray}
with $V$ the $d$-dimensional volume.
These propagators have two components, corresponding to a longitudinal part $L^{\rm diag/off}(p^2)$ and a transverse part $T^{\rm diag/off}(p^2)$:
\begin{align}
G_{\mu\nu}^{\rm diag/off}(p) = \left(\delta _{\mu\nu} - \frac{p_\mu p_\nu}{p^2}\right) T^{\rm diag/off}(p^2) + \frac{p_\mu p_\nu}{p^2}L^{\rm diag/off}(p^2).
\end{align}
Each of components may be extracted from the propagator:
\begin{align}
L^{\rm diag/off}(p^2) =& \frac{p_\mu p_\nu}{p^2}G_{\mu\nu}^{\rm diag/off}(p), \notag \\
T^{\rm diag/off}(p^2) =& \frac{1}{d-1}\left( G_{\mu\mu}^{\rm diag/off}(p) - \frac{p_\mu p_\nu}{p^2}G_{\mu\nu}^{\rm diag/off}(p) \right).
\end{align}

In MAG with U(1)$_3$ Landau gauge, in continuum limit, the diagonal gluon propagator does not have the longitudinal part $L^{\rm diag}=0$, while the off-diagonal gluon propagator have the two components.
\section{Momentum-space propagators in MAG with U(1)$_3$ Landau gauge}
\label{4}
\subsection{Gribov-Zwanziger action in MAG with U(1)$_3$ Landau gauge}
In this section, we study the momentum-space propagators in the MAG with U(1)$_3$ Landau gauge in two dimensions using SU(2) lattice QCD. First, to investigate the result from the perspective of the Gribov copies, we briefly review the behavior of the propagator obtained from the SU(2) Gribov-Zwanziger action in the Landau gauge \cite{Gribov:1977wm,Zwanziger:1989mf} and MAG \cite{Capri:2005tj,Gongyo:2013rua}. 

In the two-dimensional Landau gauge, the gluon propagator is qualitatively well described by the propagator at tree level obtained from the original Gribov-Zwanziger action, 
\begin{align}
G_{\mu \nu}(p)=\frac{p^2}{p^4+M_G^4}\left[ \delta_{\mu \nu}-\frac{p_\mu p_\nu}{p^2}\right] \label{prop_in_GZ}
\end{align}
with $M_G$ the Gribov mass, while in other dimensions, the lattice result does not coincide with the behavior of the original Gribov-Zwanziger action. This indicates that, in two dimensions, the effect of the Gribov copies is stronger than other dimensions.

In the MAG with the U(1)$_3$ Landau gauge, the Gribov-Zwanziger action has been constructed and the tree-level propagators are given by
\begin{align}
G_{\mu \nu}^{\mathrm{diag}}(p)&=\frac{p^2}{p^4+M_G^4}\left[ \delta_{\mu \nu}-\frac{p_\mu p_\nu}{p^2}\right], \notag \\
G_{\mu \nu}^{\mathrm{off}}(p)&=\frac{1}{p^2}\left[ \delta_{\mu \nu}-\frac{p_\mu p_\nu}{p^2}\right].
\end{align}
The diagonal gluon propagator behaves like the propagator in the Landau gauge given by Eq. (\ref{prop_in_GZ}), while the off-diagonal propagator shows an ordinary massless behavior. This indicates that the diagonal propagator is more greatly affected by the Gribov region than the off-diagonal propagator. 

\subsection{The diagonal propagator}
\begin{figure}[h]
\begin{center}
\includegraphics[scale=0.7]{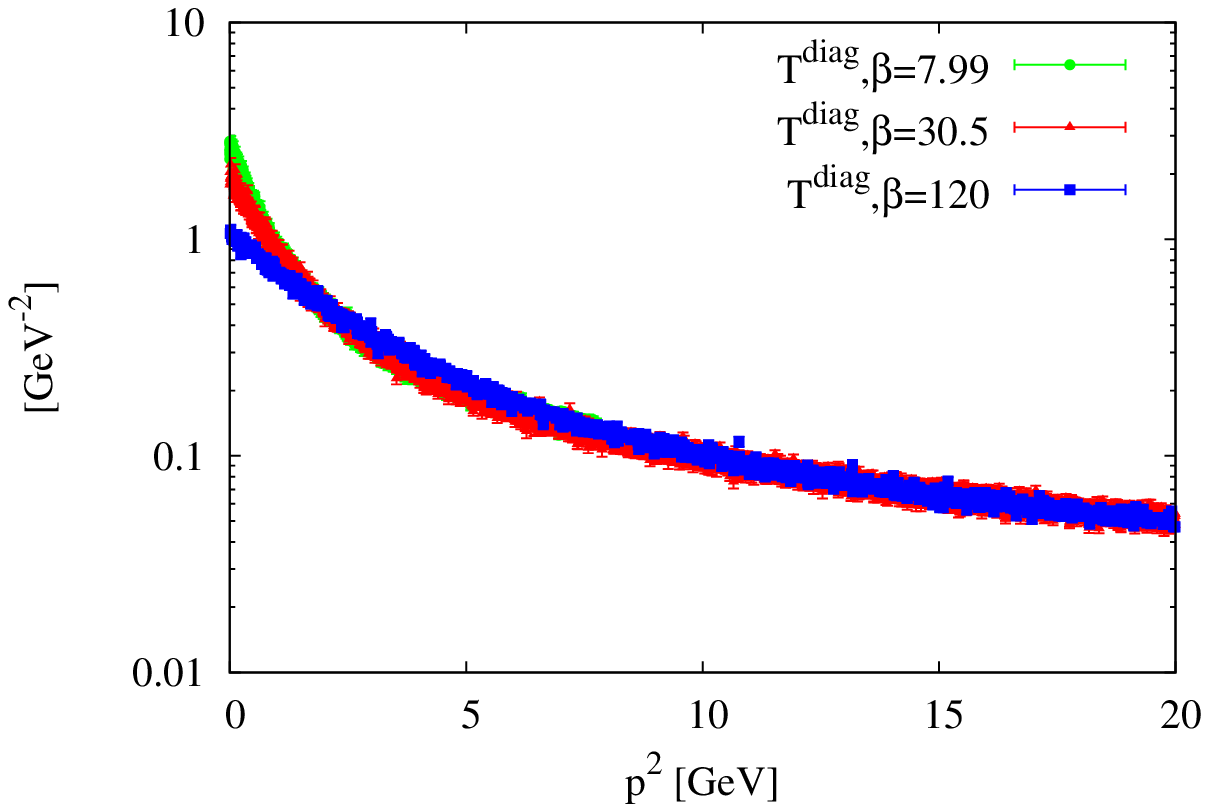}
\caption{\label{fig6}
The two-dimensional result of the diagonal gluon propagator in the MAG with the U(1)$_3$ Landau gauge fixing in momentum space. The simulation is performed at $\beta=7.99, 30.5,$ and $120$ on $256^2$ using the SU(2) lattice QCD. The gluon propagator in the infrared region decreases with increasing $\beta$.
}
\end{center}
\end{figure}

 In Fig.\ref{fig6}, we show the transverse component of the diagonal propagator on $256^2$ at $\beta =7.99,30.5,$ and $120$. Note that the longitudinal component vanishes in the continuum limit.  The lattice spacings $a$ are determined using a string tension of $(440 \mathrm{MeV})^2$. At $\beta=7.99$, $30.5$, and $120$, the lattice spacings $a$ are given by $a \simeq 0.20$ fm, $0.10$ fm, and $0.05$ fm, respectively \cite{Cucchieri:2008zx,Maas:2007uv}. We determine the renormalization factor so as to satisfy
\begin{align}
T^{\mathrm{diag}}(p^2)|_{p^2=\mu ^2}=\frac{1}{\mu ^2}
\end{align}
at the largest momentum. The result of the diagonal gluon propagator in the infrared region shows the $\beta$-dependence. Particularly, in the deep infrared region of $p^2 \le 1$GeV, the propagator is more suppressed as $\beta$ increases. This behavior is of the interest due to the following two reasons. First, if this goes to zero with increasing $\beta$, the propagator seems to be well described by the tree-level propagator in the Gribov-Zwanziger action in the MAG. Therefore this suppression may be related to the Gribov effect. Second, this behavior supports that the Abelian dominance, which means in momentum space that the diagonal propagator is enhanced in the infrared region, is not satisfied in two dimensions in contrast to four dimensions \cite{Bornyakov:2003ee,Gongyo:2013sha}.  

\begin{figure}[h]
\begin{center}
\includegraphics[scale=0.7]{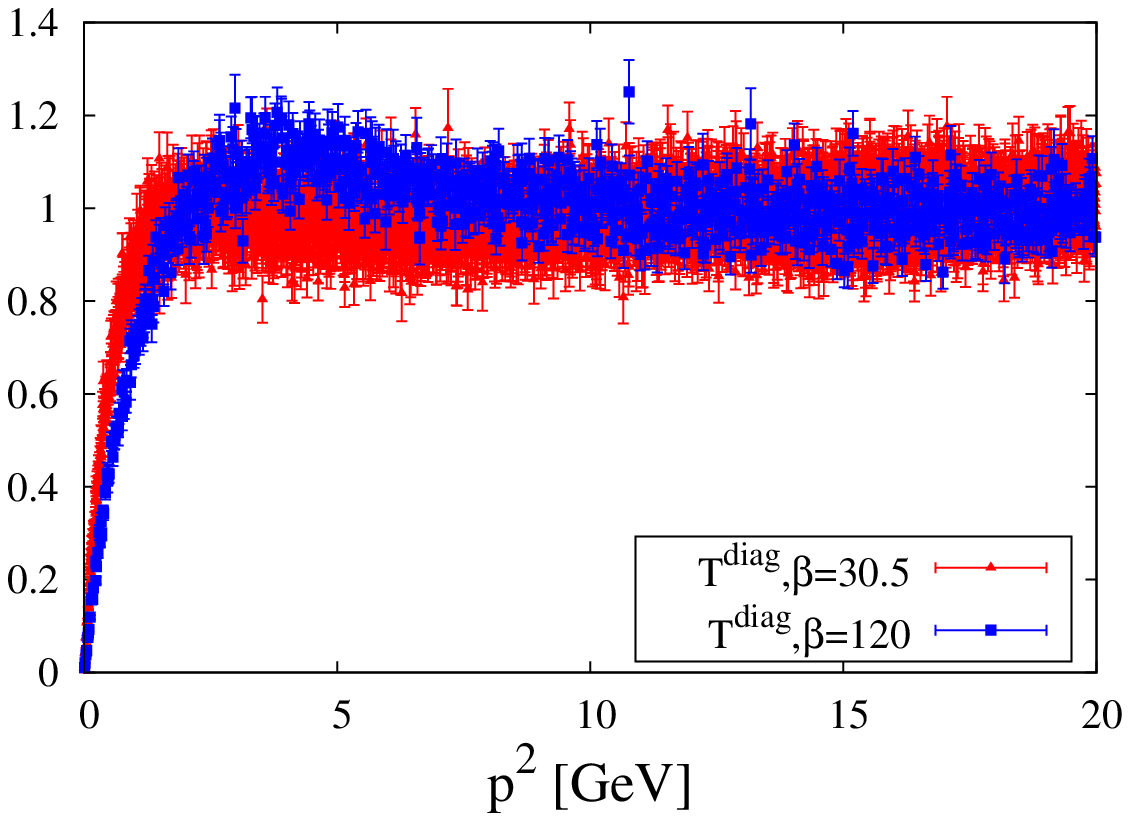}
\includegraphics[scale=0.7]{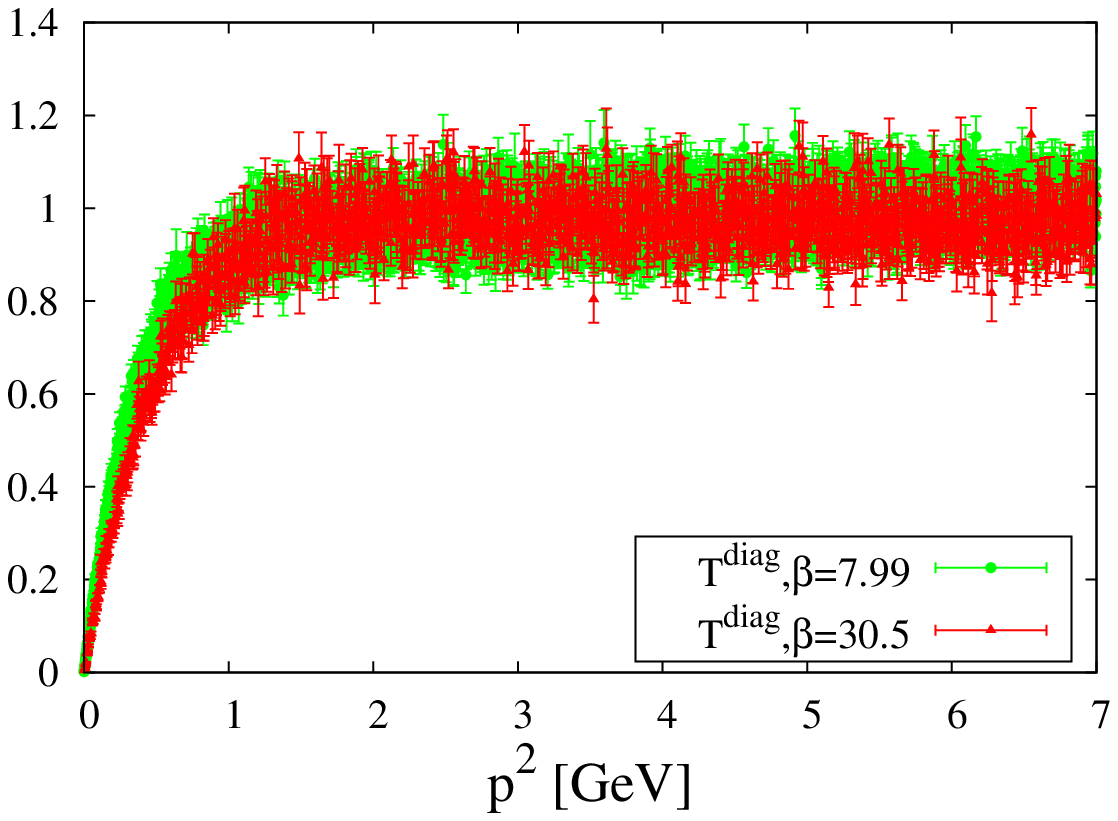}
\caption{\label{fig7}
The diagonal dressing function in the two-dimensional MAG with the U(1)$_3$ Landau gauge fixing. The simulation is performed at $\beta=7.99$, $30.5,$ and $120$ on $256^2$ using the SU(2) lattice QCD. At $\beta=120$, the dressing function has a minimum at $p^2 \simeq 4$GeV.
}
\end{center}
\end{figure}

The Gribov effect is more supported by the behavior of the gluon dressing function, $p^2T^\mathrm{diag}(p^2)$. If we consider a free massive vector field, the dressing function is given by $\frac{p^2}{p^2+m^2}$ with $m$ a mass, which shows a monotonically increasing function and approaches a constant with increasing momentum. Therefore, if the dressing function does not show an increase for some momenta, it is likely to lead to the violation of Kallen-Lehmann representation, which is related to the Gribov problem. The result for the dressing function at $\beta=7.99, 30.5,$ and $120$ is shown in Fig.\ref{fig7}. At $\beta=7.99$ and $30.5$, the dressing function does not show a sign of the violation. However, at $\beta=120$, corresponding to the finest lattice, the dressing function has a maximum at $p^2 \simeq 4$GeV. This seems the sign of the violation of the Kallen-Lehmann representation.

\begin{figure}[h]
\begin{center}
\includegraphics[scale=0.7]{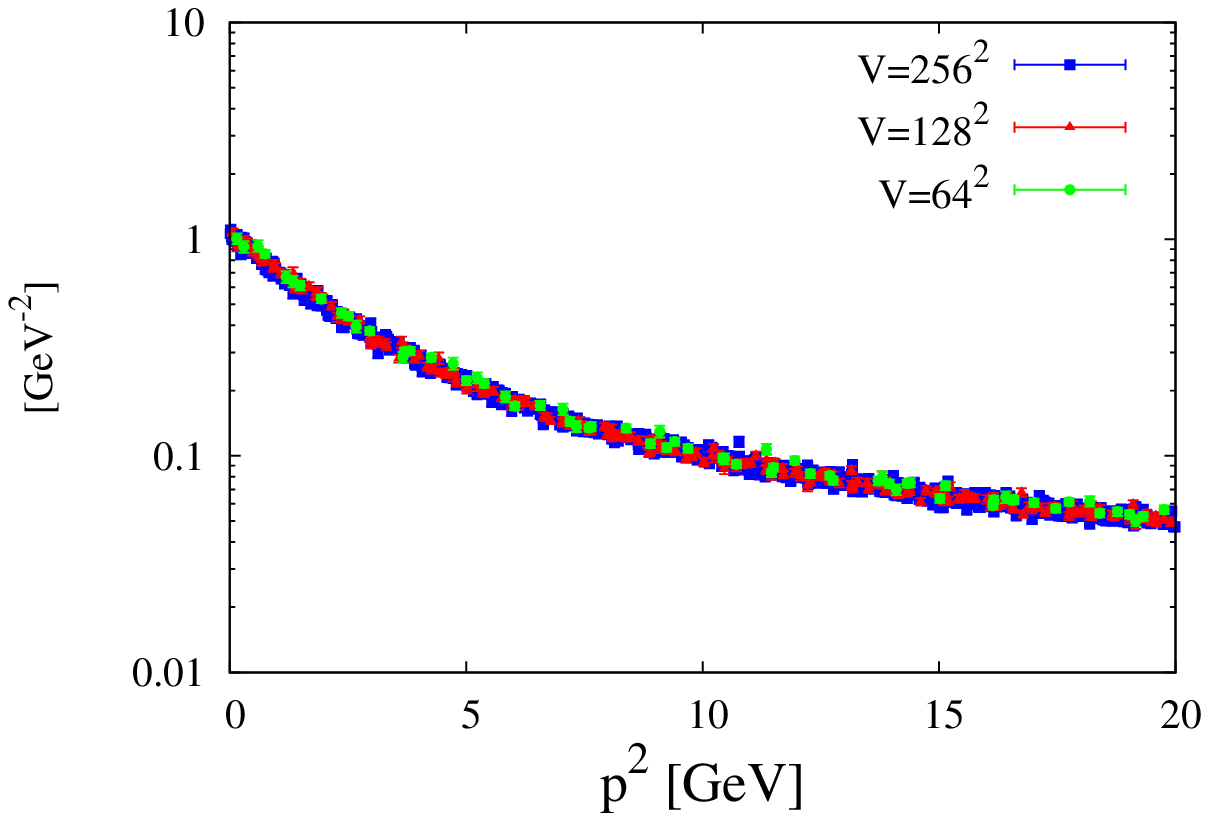}
\includegraphics[scale=0.7]{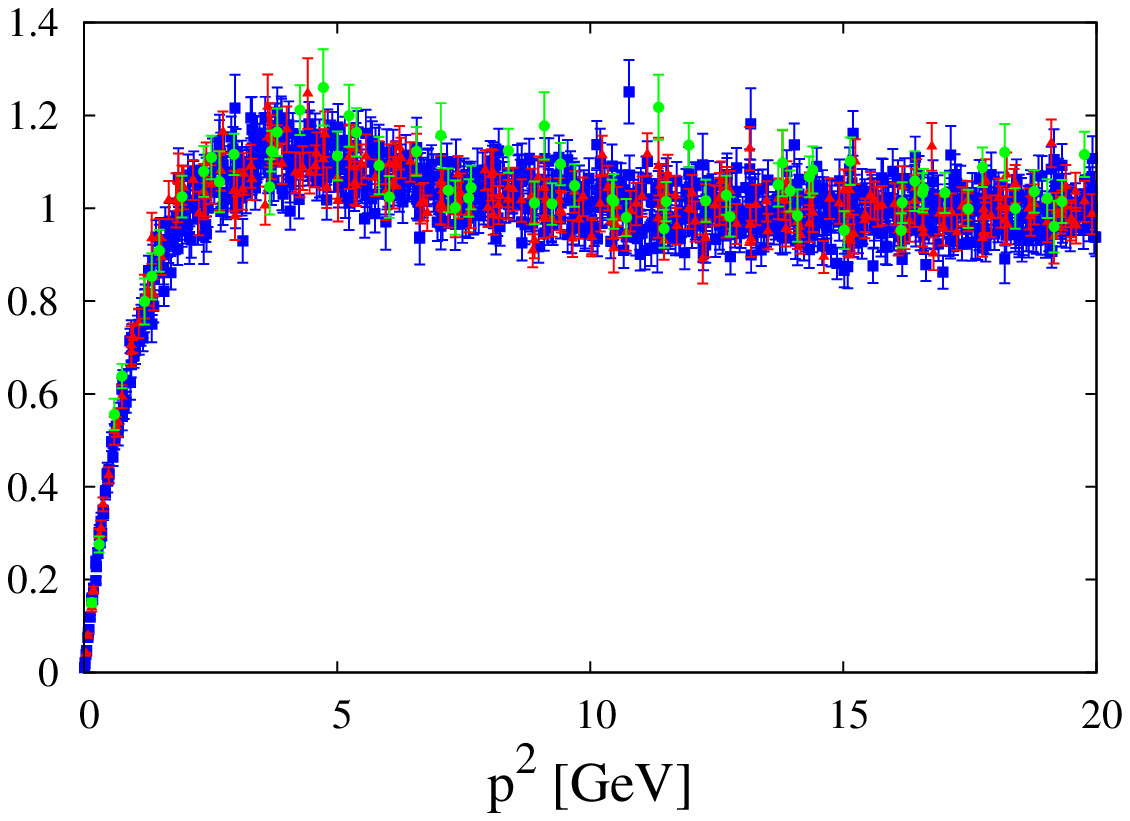}
\caption{\label{fig8}
The volume-dependence for the diagonal gluon propagator and the dressing function between $64^2$, $128^2$, and $256^2$. The finite-volume effect is not found.
}
\end{center}
\end{figure}

Finally we study the diagonal gluon propagator and the dressing function for three different volumes, $64^2$, $128^2$, and $256^2$, at $\beta=120$ in Fig.\ref{fig8}. Even in the infrared region, the finite-volume effects are hardly visible. Therefore the diagonal gluon propagator does not seem to be enhanced even in infinite volume limit and this behavior also supports that the Abelian dominance is not satisfied in two dimensions. 

\subsection{The off-diagonal propagator}
\begin{figure}[h]
\begin{center}
\includegraphics[scale=0.7]{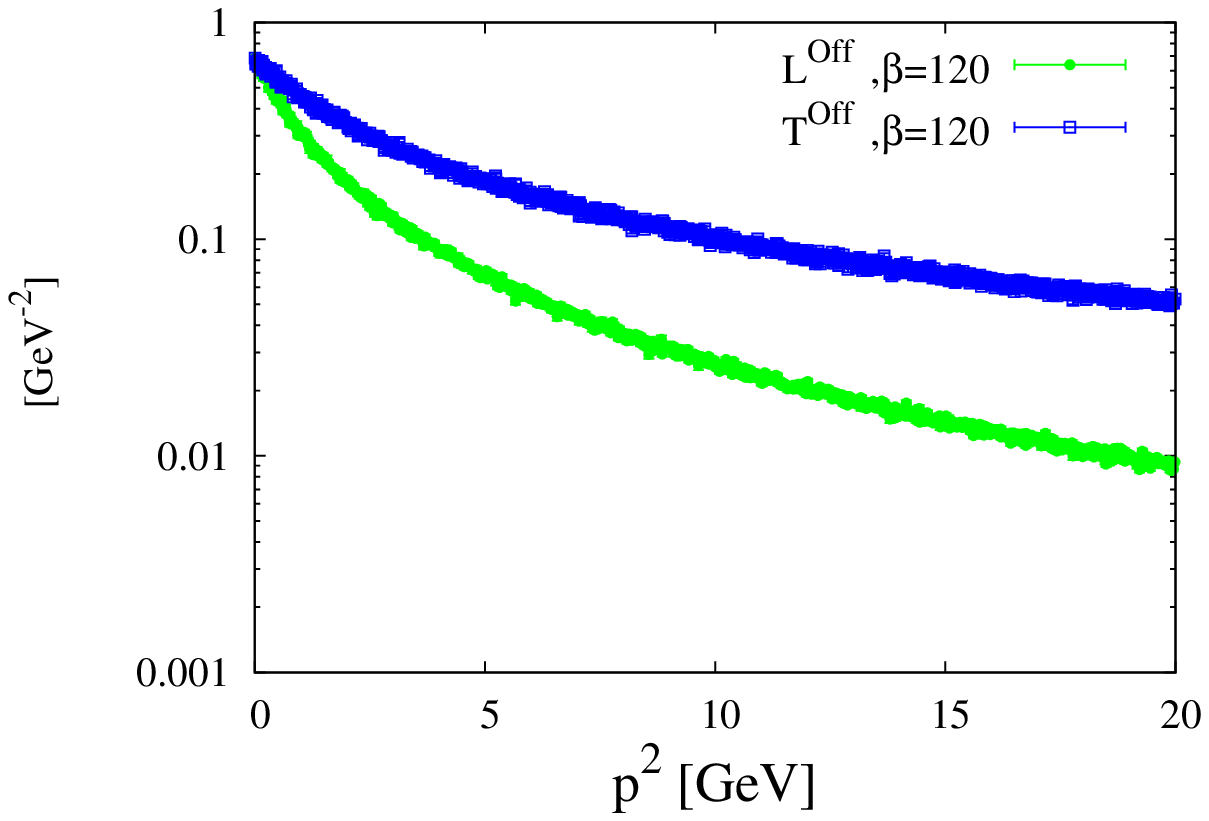}
\caption{\label{fig9}
The transverse and longitudinal components of the off-diagonal gluon propagator in the two-dimensional MAG at $\beta=120$ on $256^2$.
}
\end{center}
\end{figure}

Next, we study the off-diagonal propagator, which has diagonal and off-diagonal components as defined in Sec.\ref{2}. The renormalization factor is determined by the tree-level transverse component of the off-diagonal propagator at the largest momentum as done in the diagonal propagator. The result of the transverse and longitudinal components at $\beta=120$ on $256^2$ is shown in Fig.\ref{fig9}. The longitudinal component decreases faster than the off-diagonal component with increasing momentum in accordance with the tree-level result which shows that the longitudinal component vanishes.  

\begin{figure}[h]
\begin{center}
\includegraphics[scale=0.7]{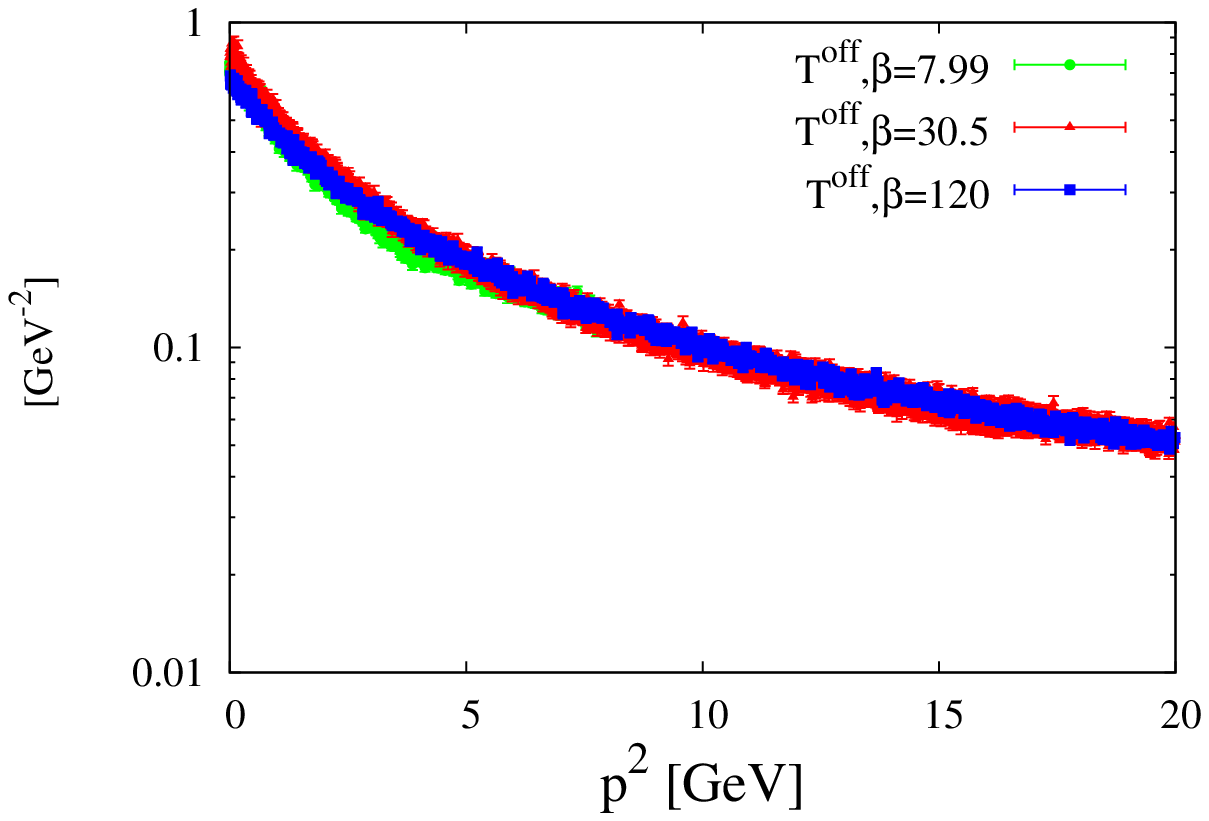}
\includegraphics[scale=0.7]{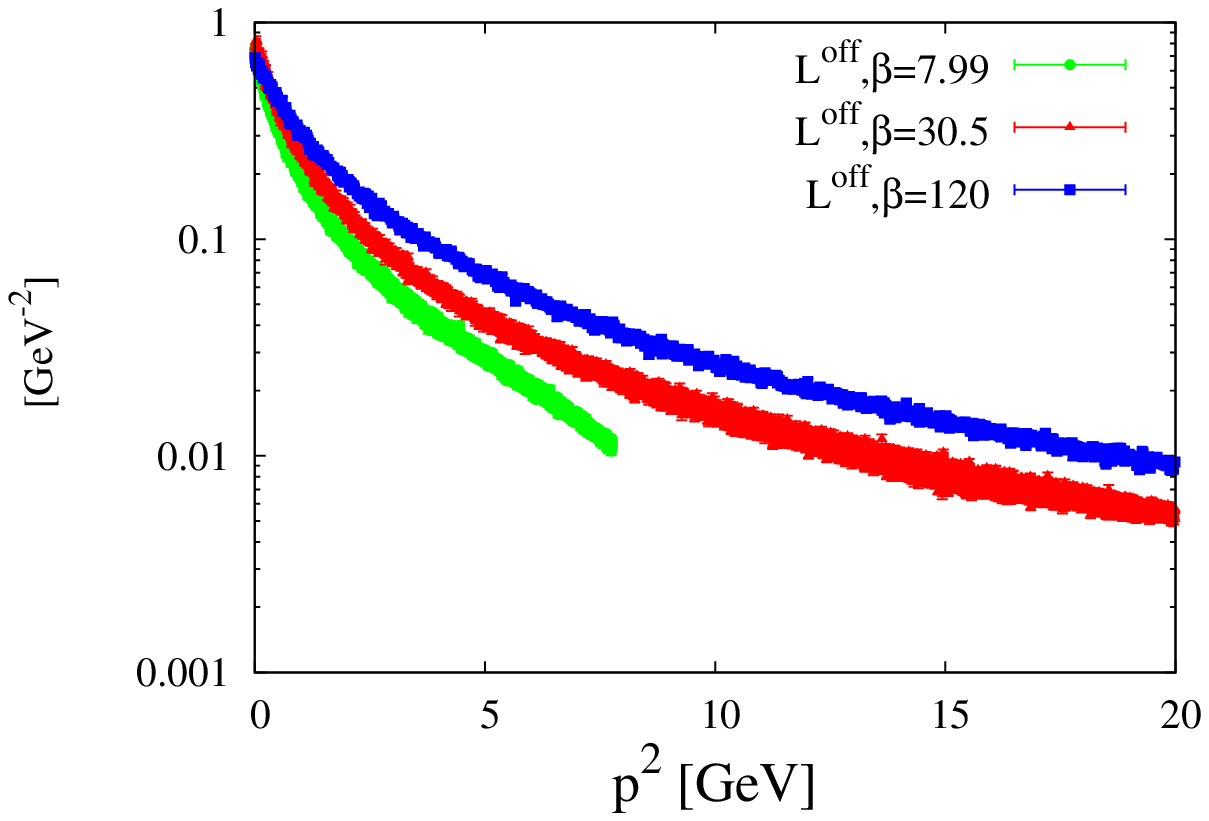}
\caption{\label{fig10}
The transverse component (top) and longitudinal component (bottom) of the off-diagonal gluon propagator in the two-dimensional MAG at $\beta=7.99,30.5,$ and $120$ on $256^2$.
}
\end{center}
\end{figure}

In Fig.\ref{fig10}, we also show the $\beta$-dependence between $\beta=7.99, 30.5,$ and $120$. The $\beta$-dependence for the transverse components is hardly visible compared to the diagonal propagator. On the other hand, the longitudinal component is largely influenced by $\beta$. Although the longitudinal component seems to be enhanced with increasing $\beta$, this behaves more singular than $1/p^2$ and thus approaches zero faster than the transverse component even at $\beta=120$ as in Fig. \ref{fig9}.

\begin{figure}[h]
\begin{center}
\includegraphics[scale=0.7]{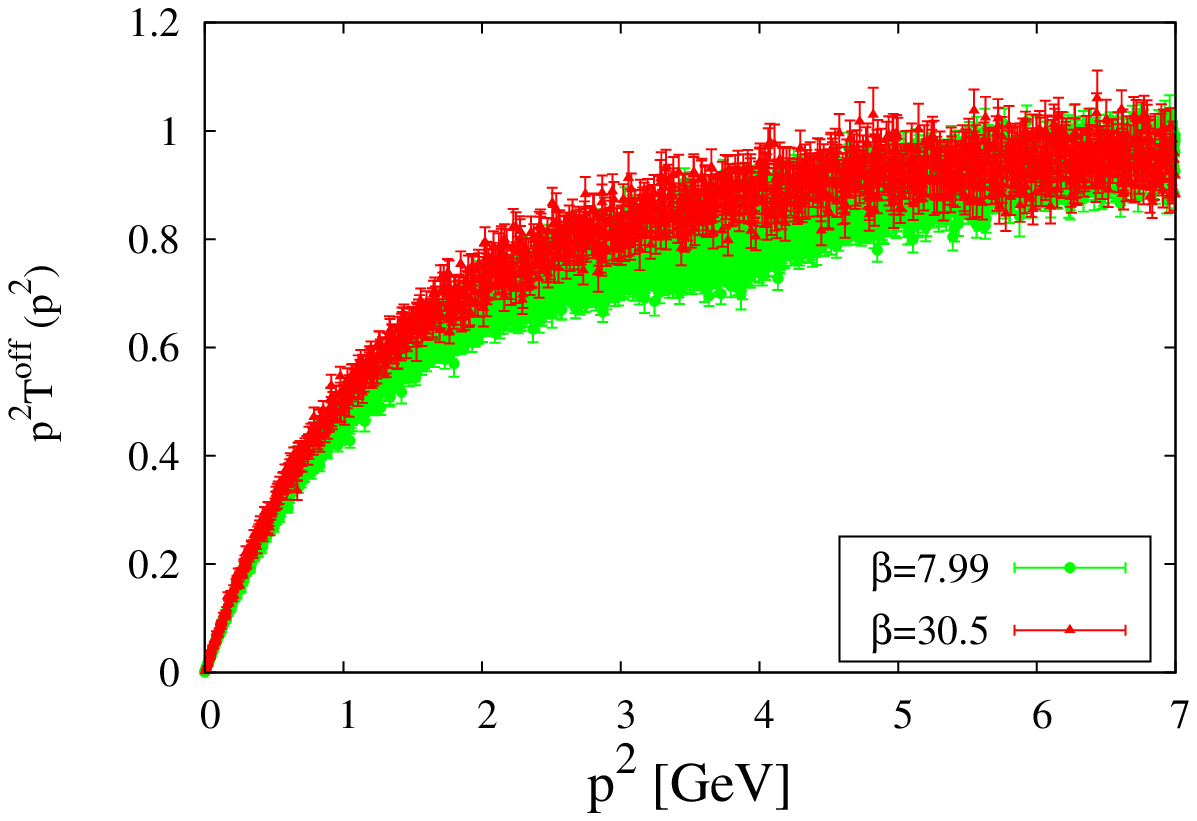}
\includegraphics[scale=0.7]{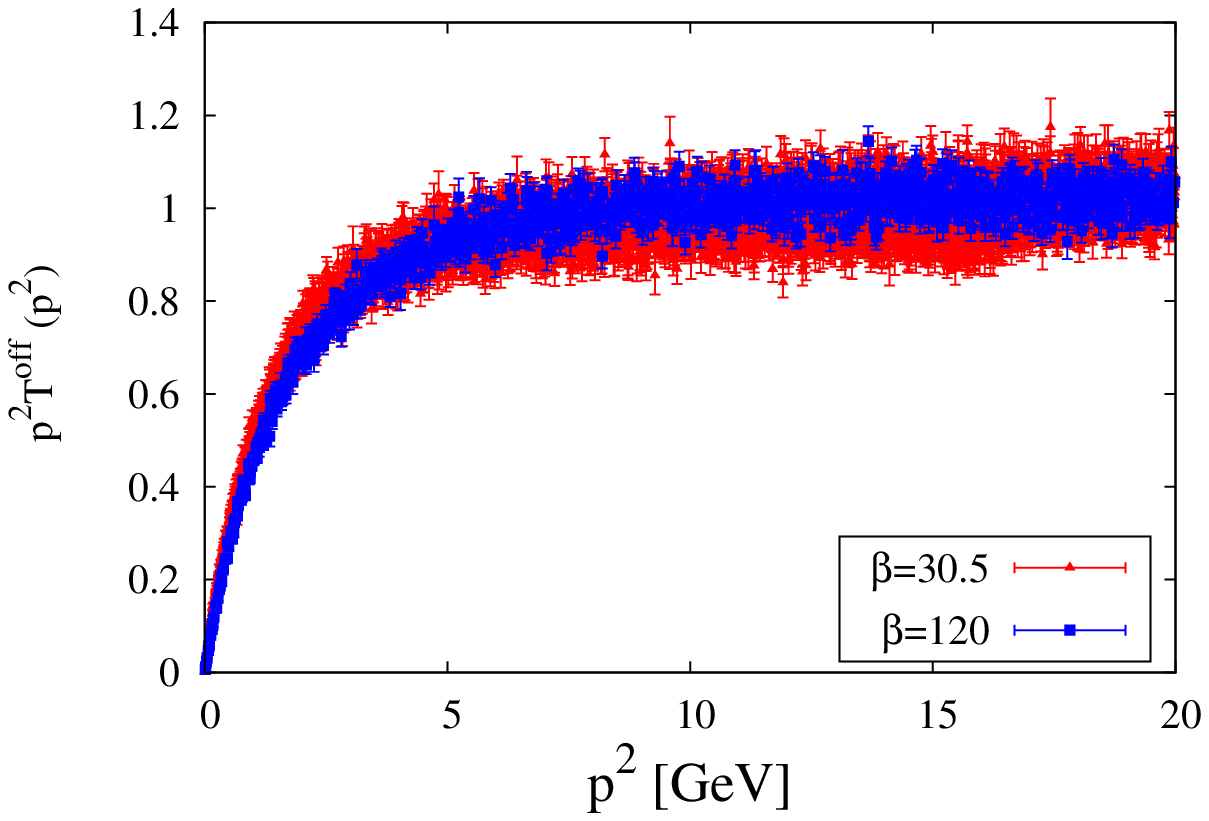}
\caption{\label{fig11}
The off-diagonal dressing function of the transverse component in the two-dimensional MAG at $\beta=7.99,30.5$, and $120$ on $256^2$.
}
\end{center}
\end{figure}

Furthermore, we show the gluon dressing function of the transverse component in Fig.\ref{fig11}. The dressing function increases monotonically and approaches 1 with increasing momentum, and thus we do not find the violation of the Kallen-Lehmann representation which is found in the diagonal dressing function. The characteristic $\beta-$dependence between $\beta=7.99,30.5$, and $120$ is also not visible. Thus we may fit the off-diagonal gluon propagator of the transverse component by the functional form, $Z/(p^2+m^2)$ with a mass parameter $m$ and a renormalization constant $Z$ as shown in Table.\ref{TableII}. In fact, the fit result shows that the propagator is well described by the functional form and the effective mass is estimated as $m_\mathrm{off} \simeq 1$GeV.
\begin{table}[h]
\caption{Summary table of conditions and results using SU(2) lattice QCD in two dimensions. 
The off-diagonal gluon mass $m_{\rm off}$ is estimated by using the functional form $Z/(p^2+m^2)$ with a mass parameter $m$ and a renormalization constant Z in the region of $p^2 \le p^2_\mathrm{max}$ which is the largest momentum.
}
\label{TableII}
\begin{center} 
\begin{tabular}{ccccc}
\hline
\hline
 $\beta$     & $a[{\rm fm}]$ &    $m_{\rm off} [{\rm GeV}] $ &$Z$ &$\chi^2/N$\\
\hline
    ~~~7.99~~~  & 0.2&1.21 & 20.5 & 2.1\\
    ~~~30.5~~~     &  0.1&1.08 & 17.5 & 1.8\\
     ~~~120~~~     & 0.05 &1.09  & 16.8 & 2.4\\
\hline
\hline
\end{tabular}
\end{center} 
\end{table}

The difference between the transverse components of the diagonal propagator and the off-diagonal propagator is of interest from the Gribov problem. The fact that the diagonal dressing function has a maximum, and the off-diagonal dressing function of the transverse component shows a monotonically increasing function indicates that the diagonal gluon is more influenced by the Gribov region than the off-diagonal gluons. This is in agreement with the behavior of the gluon propagator obtained from the Gribov-Zwanziger action in the MAG, though the lattice result at $\beta=120$ does not behave like the tree-level propagator.

\begin{figure}[h]
\begin{center}
\includegraphics[scale=0.7]{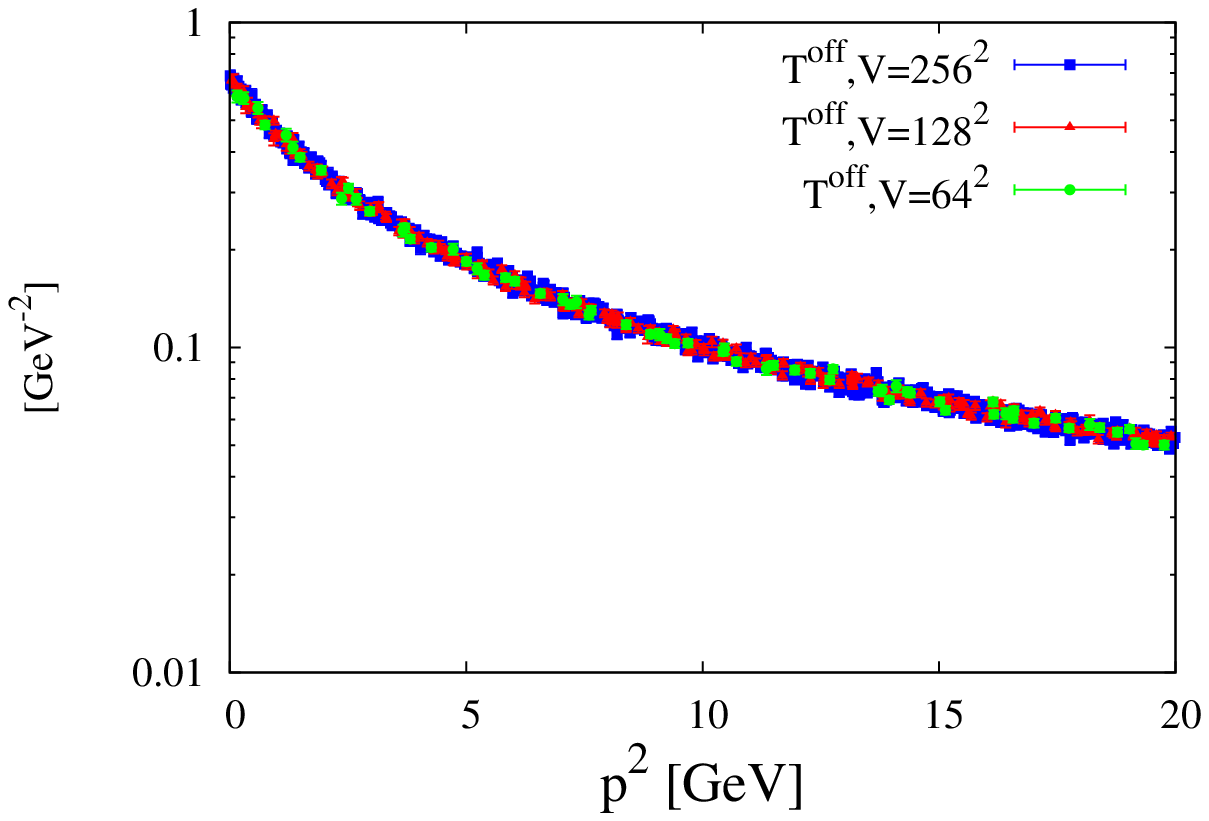}
\includegraphics[scale=0.7]{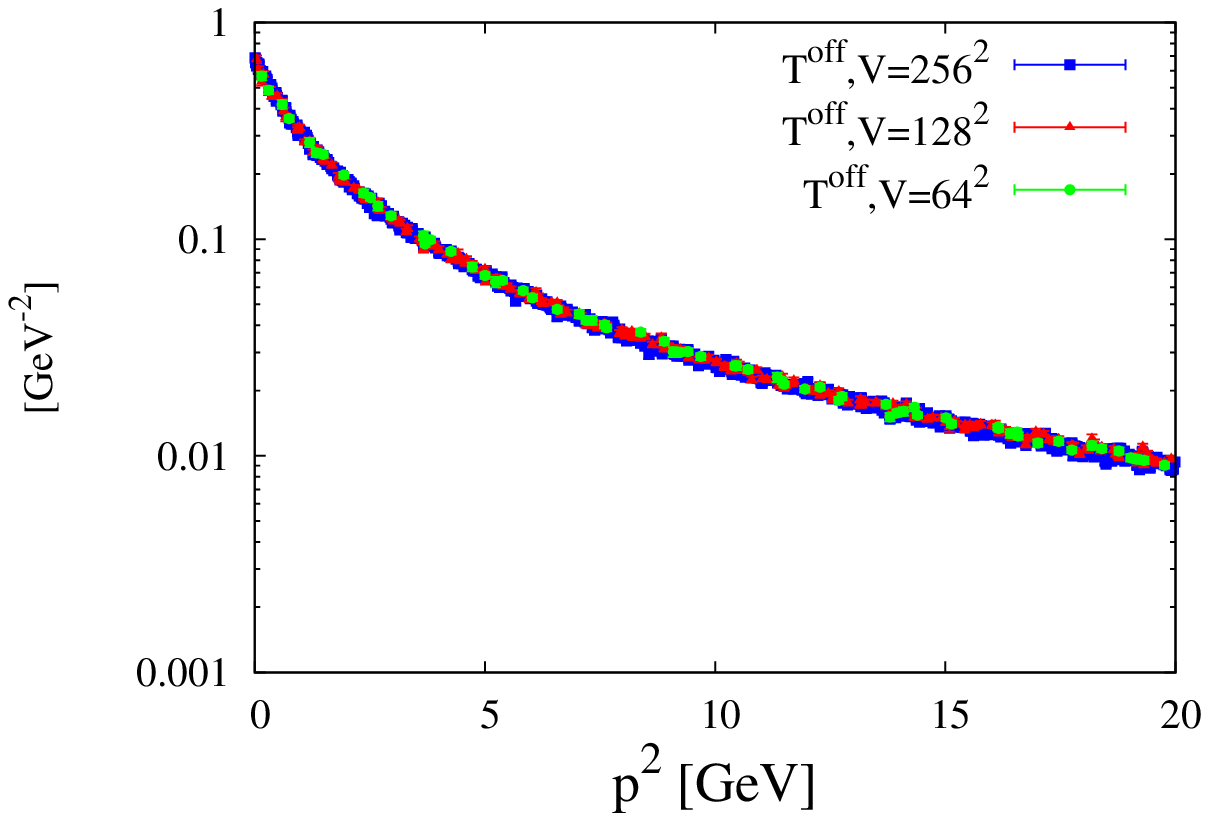}
\caption{\label{fig12}
The volume-dependence for the transverse component (top) and longitudinal component (bottom) of the off-diagonal gluon propagator in the two-dimensional MAG at $\beta=120$ on $62^2,128^2$, and $256^2$.
}
\end{center}
\end{figure}

Finally, we  show the volume-dependence for the off-diagonal gluon propagators of the longitudinal and transverse components between $64^2, 128^2,$ and $256^2$ in Fig.\ref{fig12}. There seems to be almost no volume-dependence between them.

\section{SU(2) gluon propagators in MAG with U(1)$_3$ Landau gauge in coordinate space}
\label{3}
\subsection{Two-dimensional diagonal and off-diagonal propagators}
\begin{figure}[h]
\begin{center}
\includegraphics[scale=0.7]{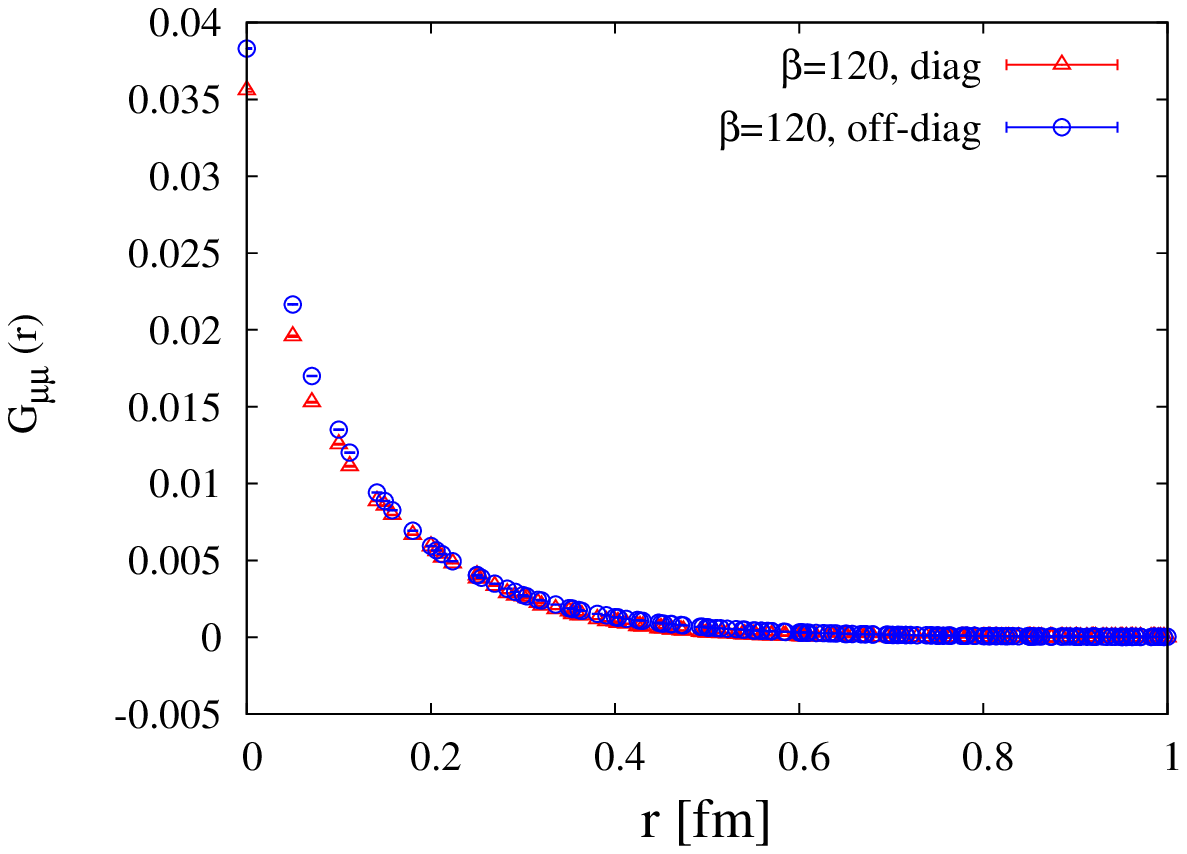}
\includegraphics[scale=0.7]{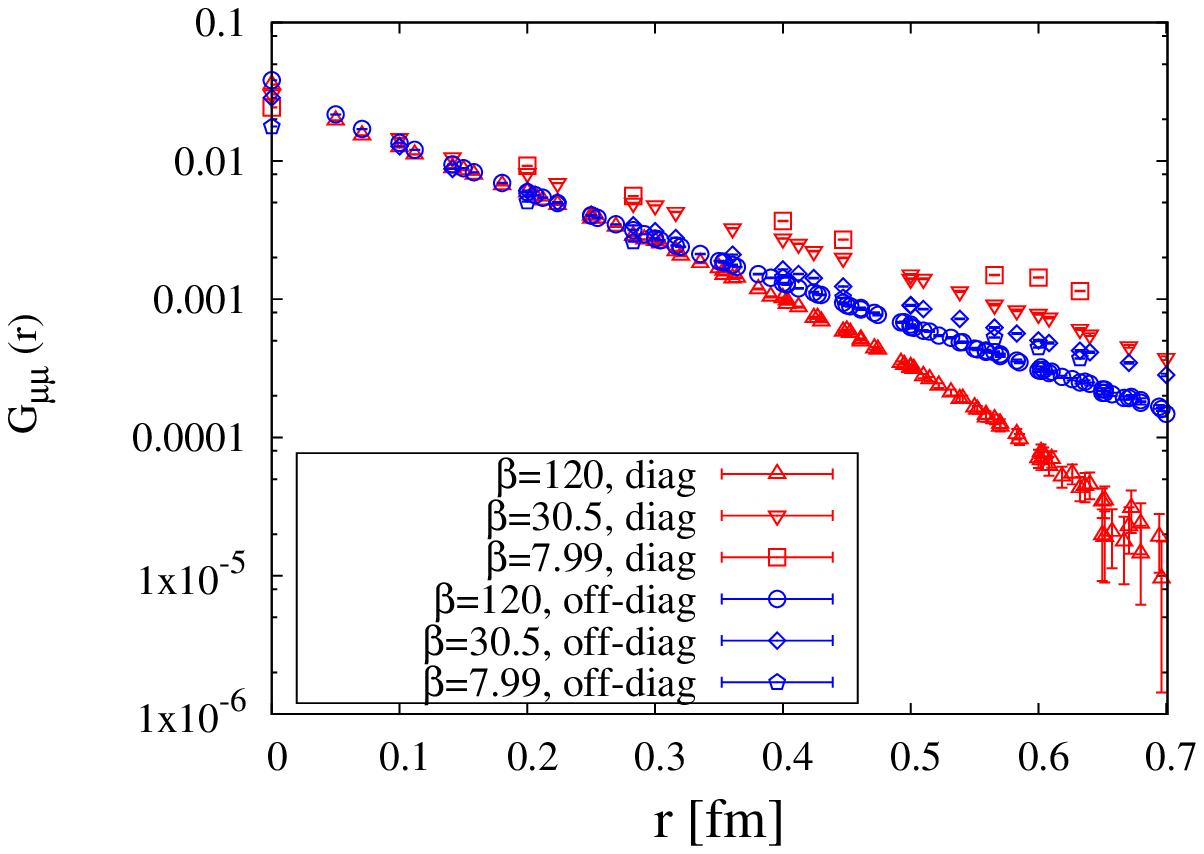}
\caption{\label{fig1}
The two-dimensional SU(2) lattice QCD results of gluon propagators $G_{\mu\mu}^{\rm diag}(r)$ and $G_{\mu\mu}^{\rm off}(r)$ (top), and their logarithmic plots (bottom) as the function of $r\equiv \sqrt{(x_\mu -y_\mu)^2}$ in MAG with the U(1)$_3$ Landau gauge fixing in the physical unit. The Monte Carlo simulation is performed on $256^2$ lattice with $\beta = 7.99,30.5,$ and $120$. 
}
\end{center}
\end{figure}

In this section, we study the two-dimensional gluon propagators in the MAG in coordinate space. The diagonal and off-diagonal gluon propagators and their logarithms in the MAG at $\beta=7.99$, $30.5,$ and $120$ on $256^2$ are shown in Fig.\ref{fig1}. The renormalization factors we determined in Sec. \ref{4} are used.

If the Abelian dominance holds even in two dimensions, the off-diagonal gluon propagator is largely reduced and the diagonal gluon propagator takes a large value at long distance \cite{Gongyo:2012jb,Gongyo:2013sha}. However, we do not find the remarkable difference between the diagonal propagator and the off-diagonal propagator at $\beta =120$ in the top panel of Fig.\ref{fig1}.
Furthermore, the logarithm of the diagonal propagator at $\beta=120$ decreases more largely than that of the off-diagonal propagator in the region of $r\gsim 0.4$ as shown in the bottom panel of Fig.\ref{fig1}. 
At long distance, the diagonal propagator largely depends on $\beta$ and is suppressed with increasing $\beta$. This behavior corresponds to the suppression of the momentum-space propagator in the infrared region as shown in Fig.\ref{fig7}. In particular, the rapid decrease in the region of $r\gsim 0.4$ at $\beta=120$ seems to be related to the presence of a maximum in the dressing function as shown in Fig.\ref{fig8}. The decrease is also the sign of the effect of the Gribov copies. In fact, the gluon propagator in the Landau gauge in two dimensions shows a similar rapid decrease and takes the negative values, and thus this would vanish at zero momentum.

\subsection{The propagators at zero-spatial momentum}
Next, we study the propagators at zero-spatial-momentum,
\begin{align}
D^0_{\mu \nu}(t) =  \frac{1}{3V}\sum_{a,x_1,y_1}\left< A_\mu ^a (x_1,t) A_\nu ^a (y_1,0)\right>
\end{align}
with the two dimensional volume $V$. Note that the spectral function is given by the inverse Laplace transformation of the propagator and thus we may investigate the behavior of the spectral function using the gluon propagator. The gluon propagator at zero-spatial-momentum has been studied from the perspective of the positive violation of the spectral function \cite{Mandula:1987rh,Mandula:1999nj,Gongyo:2012jb,Gongyo:2013sha}. The evidence of the violation is obtained by the behavior of the logarithm of the propagator which shows concave downward.

The longitudinal and transverse parts of the propagator at zero-spatial momentum, $D^0 _L(t)$ and $ D^0 _T(t) $, are extracted from the time-time component and space-space component, respectively \cite{Gongyo:2012jb,Gongyo:2013sha}:
\begin{align}
D^0_{\mu \nu}(t)=\delta^{\mu 1 \nu 1}D^0_T(t) +\delta^{\mu 4 \nu 4}D^0_L(t) .
\end{align}

\begin{figure}[h]
\begin{center}
\includegraphics[scale=0.7]{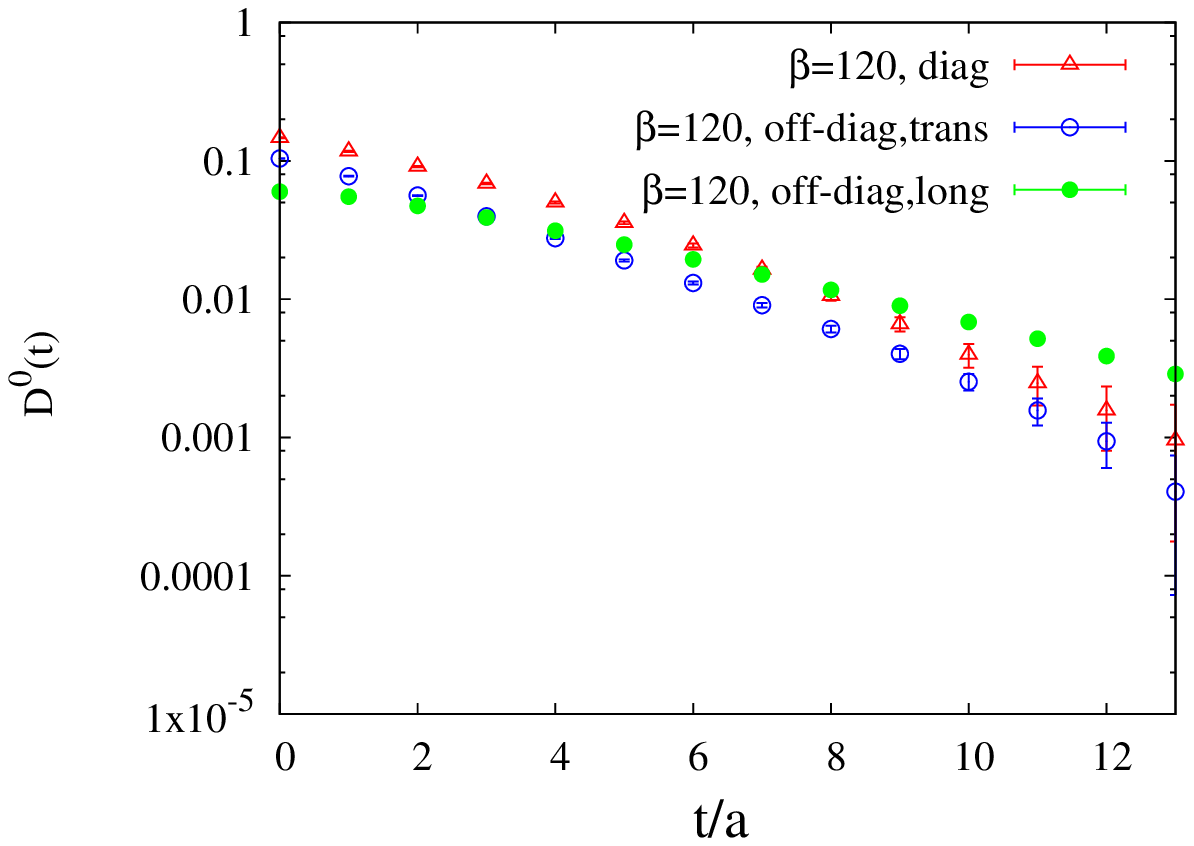}
\caption{\label{fig3}
The logarithmic plots of the diagonal and off-diagonal gluon propagators at zero-spatial-momentum in the region of $t/a=0-13$. The Monte Carlo simulation is performed on $256^2$ lattice at $\beta = 120$.
}
\end{center}
\end{figure}

In Fig.\ref{fig3}, we show the logarithms of the (transverse) diagonal propagator and the transverse and longitudinal off-diagonal propagator at zero-spatial-momentum at $\beta=120$. The result shows that all of these propagators have a curve which is concave downward and thus this leads to the violation of the positivity for the spectral functions.  

\subsection{mass estimation and comparison with four dimensional propagators}
\label{AD}
 Finally, to see qualitatively whether the Abelian dominance is supported or not in two dimensions, we estimate the diagonal and off-diagonal effective masses using the Proca formalism as shown in Appendix \ref{ap1}, and compare the result with the four dimensional result. According to the Proca formalism, in two dimensions, the propagator behaves like $G_{\mu\mu}(r) \sim e^{-Mr}/r^{1/2}$ with $M$ the effective mass in the infrared region. Therefore we obtain the effective masses from the slope of the logarithms of $r^{1/2}G_{\mu\mu}^{\rm diag} (r)$ and $r^{1/2}G_{\mu\mu}^{\rm off} (r)$ in two dimensions. On  the other hand, in four dimensions, the propagator behaves like $G_{\mu\mu}(r) \sim e^{-Mr}/r^{3/2}$ in the infrared region and the effective masses are obtained from the slope of the logarithms of $r^{3/2}G_{\mu\mu}^{\rm diag} (r)$ and $r^{3/2}G_{\mu\mu}^{\rm off} (r)$. 

\begin{figure}[h]
\begin{center}
\includegraphics[scale=0.7]{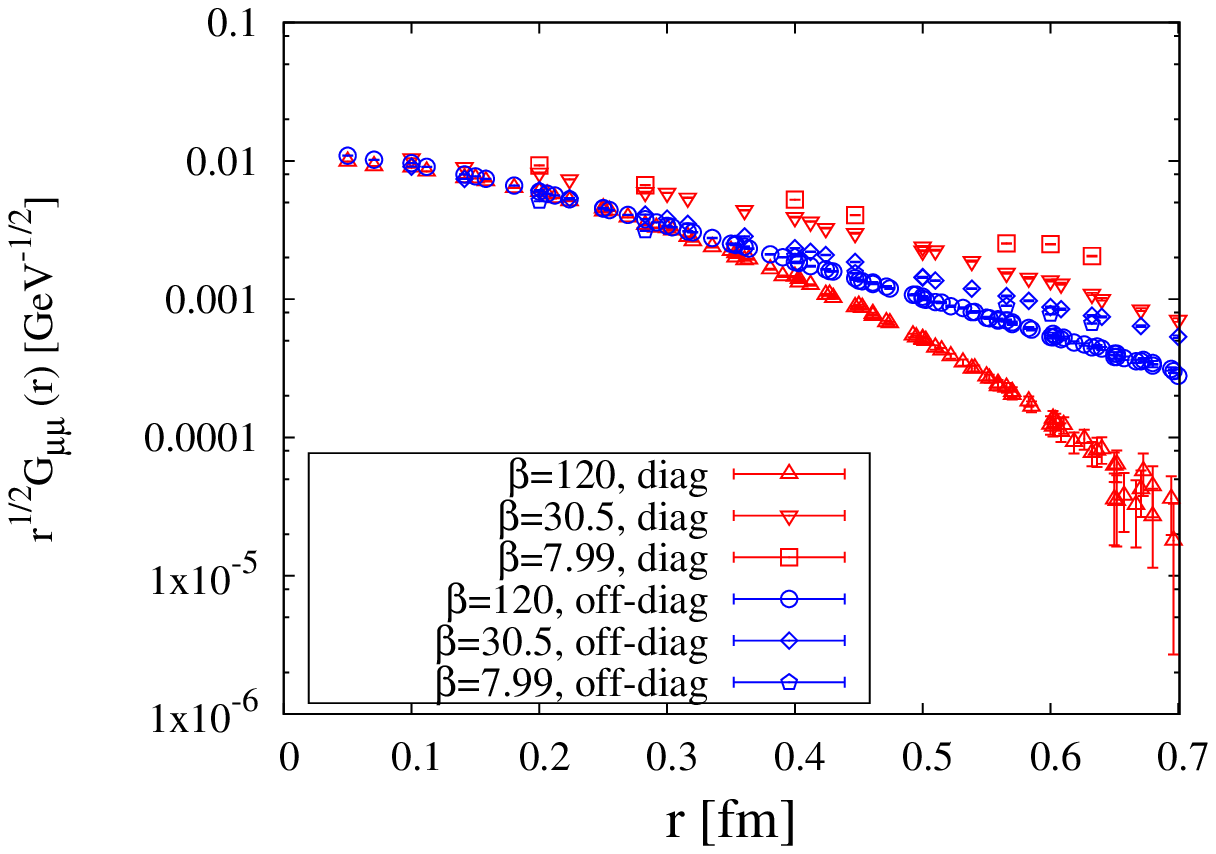}
\caption{\label{fig4} 
The two-dimensional result of logarithmic plots of $r^{1/2}G_{\mu\mu}^{\rm off} (r)$ and $r^{1/2}G_{\mu\mu}^{{\rm diag}} (r)$ in the MAG with the U(1)$_3$ Landau gauge fixing. The simulation is performed at $\beta=7.99, 30.5,$ and $120$ using the SU(2) lattice QCD.
}
\end{center}
\end{figure}

In Fig.\ref{fig4}, we show the logarithmic plots of $r^{1/2}G_{\mu\mu}^{\rm diag} (r)$ and $r^{1/2}G_{\mu\mu}^{\rm off} (r)$ at $\beta=7.99,30.5,$ and $120$ on $256^2$ in two dimensions. In Table.\ref{TableI} we summarize the effective diagonal and off-diagonal gluon masses, $M_\mathrm{diag}$ and $M_\mathrm{off}$, obtained by fitting the slope in the region of $r=0.4-0.8$ fm. At $\beta=7.99$ and $30.5$, the diagonal gluon mass is similar to the off-diagonal mass. At $\beta =120$, closest to a continuum limit, the diagonal gluon mass is larger than the off-diagonal mass. The behavior corresponds to a drastic decrease in the diagonal gluon propagator at long distance and thus might be related to the Gribov problem. The result supports that the Abelian dominance is not found, and the Gribov effects for the diagonal gluon are found in two dimensions.

\begin{figure}[h]
\begin{center}
\includegraphics[scale=0.7]{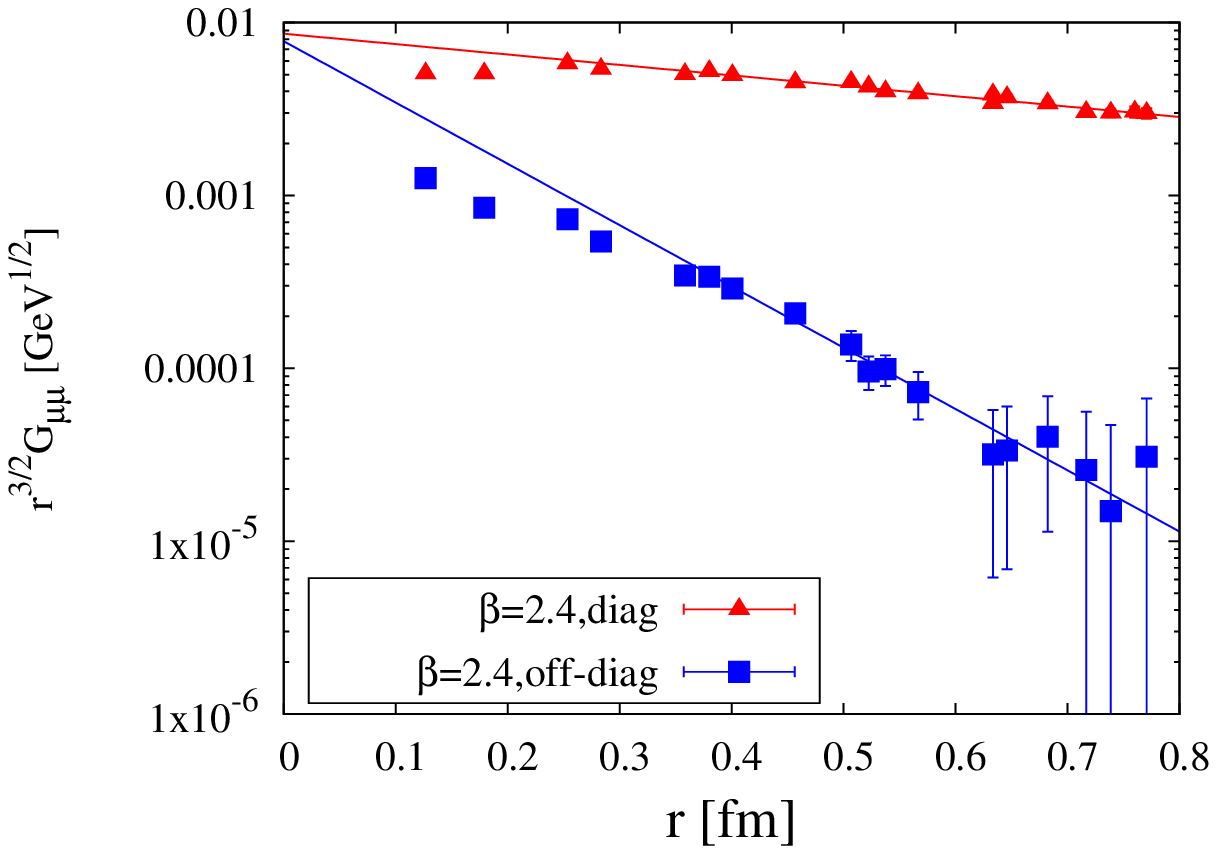}
\caption{\label{fig5}
The four-dimensional result of logarithmic plots of $r^{3/2}G_{\mu\mu}^{\rm off} (r)$ and $r^{3/2}G_{\mu\mu}^{{\rm diag}} (r)$ in the MAG with the U(1)$_3$ Landau gauge fixing. The simulation is performed at $\beta=2.4$ on $16^4$ using the SU(2) lattice QCD.
}
\end{center}
\end{figure}

To compare the result with the  four-dimensional behavior, we also show the four-dimensional gluon propagators. The logarithmic plots of $r^{3/2}G_{\mu\mu}^{\rm diag} (r)$ and $r^{3/2}G_{\mu\mu}^{\rm off} (r)$ at $\beta=2.6$ on $16^2$ are shown in Fig.\ref{fig5} and the effective masses are summarized in Table.\ref{TableI}. The diagonal propagation seems to be dominant at long distance in contrast to two dimensions. In fact, the off-diagonal gluon has a large effective mass, $M_{\mathrm{off}}\simeq 1.6 \mathrm{GeV}$, while the diagonal gluon has a small effective mass, $M_{\mathrm{diag}} \simeq 0.3 \mathrm{GeV}$. Therefore only the diagonal gluon propagates widely and the infrared phenomena seem to be dominated by the diagonal gluon.

The diagonal propagator in MAG depends largely on the dimensionality, similar to the Landau gauge, while the off-diagonal gluon propagator seems to show a similar behavior among two and four dimensions. The reason why the Abelian dominance is not found in two dimensions might be the disappearance of monopole. In four dimensional MAG, the monopole is likely to appear in the diagonal part and it supports the infrared phenomena \cite{Kronfeld:1987vd,Kronfeld:1987ri,Brandstater:1991sn,Stack:1994wm,Miyamura:1995xn,Woloshyn:1994rv}. However, in two dimensions, the monopole cannot appear because of the topology and thus the Abelian dominance is not found. 

In addition, the dependence for the dimensionality for the diagonal gluon is likely to be related to the Gribov problem. In the Landau gauge, the two-dimensional gluon propagator is well described by the propagator obtained by the original Gribov-Zwanziger action in contrast to other dimensions. This suggest that the difference between the dimensions is related to the influence of the Gribov copies for the dimensionality. Therefore, the difference between two- and four-dimensional diagonal gluon in the MAG also indicates the effect of the Gribov copies and no remarkable difference between two- and four-dimensional off-diagonal propagator indicates that the effect is weaker for the off-diagonal gluons than the diagonal gluons, consistent with the Gribov-Zwanziger action in the MAG.
\begin{table}[h]

\caption{Summary table of conditions and results in SU(2) lattice QCD in two and four dimensions. 
The off-diagonal gluon mass $M_{\rm off}$ and the diagonal gluon mass $M_{\rm diag}$ are 
obtained from the slope analysis of $r^{1/2}G_{\mu\mu}^{\rm off} (r)$ and $r^{1/2}G_{\mu\mu}^{\rm diag} (r)$ in two dimensions and $r^{3/2}G_{\mu\mu}^{\rm off} (r)$ and $r^{3/2}G_{\mu\mu}^{\rm diag} (r)$ in four dimensions for $r=0.4-0.8{\rm fm}$.
}
\label{TableI}
\begin{center} 
\begin{tabular}{ccccc}
\hline
\hline
  lattice size   & $\beta$     & $a[{\rm fm}]$ &   $M_{\rm diag} [{\rm GeV}] $ & $M_{\rm off} [{\rm GeV}] $  \\
\hline
$256^2$                 &    ~~~7.99~~~  & 0.20 &  0.8  & 0.9\\
                 &    ~~~30.5~~~     &  0.10 & 1.1 & 1.0\\
                 &    ~~~120~~~     &  0.05 & 2.2 & 1.2\\
$16^4$                 &    ~~~2.4~~~  & 0.13 &  0.28  & 1.6\\
\hline
\hline
\end{tabular}
\end{center} 
\end{table}

\section{Summary and concluding remarks}
\label{5}
We have investigated the gluon propagators in the SU(2) MAG with the U(1)$_3$ Landau gauge fixing in momentum space and coordinate space. The Monte Carlo simulation is performed on the $64^2$, $128^2,$ and $256^2$ lattice at $\beta = 7.99, 30.5,$ and $120$ at the quenched level.

In momentum space, we have calculated the transverse component of the diagonal gluon propagator, and the transverse and longitudinal components of the off-diagonal gluon propagator in two dimensions. In the infrared region, the transverse component in the diagonal propagator is suppressed as $\beta$ increases, while the transverse component in the off-diagonal propagator shows no $\beta$-dependence. Furthermore, at $\beta=120$, the dressing function of the transverse diagonal propagator has a maximum at $p^2 \simeq 4$GeV and that of the transverse off-diagonal propagator shows a monotonically increasing and approaches a finite value. These behaviors support that the Abelian dominance is not found in two dimensions, and the effect of the Gribov copies just appear in the diagonal propagator in accordance with the behavior of the Gribov-Zwanziger action in the MAG. However, we have not found that the tree-level propagator of the action in the MAG coincides with our result in contrast to the two dimensional Landau gauge. As for the longitudinal component of the off-diagonal propagator, the behavior shows more singular than $p^2$ and thus this would vanish in the ultraviolet region, though the $\beta$-dependence has been found.

In coordinate space, we have shown the scalar combinations of the diagonal and off-diagonal propagators $G_{\mu\mu} (r)$. At long distance, the diagonal gluon propagator is suppressed with $\beta$ increasing. Particularly, at $\beta=120$, the diagonal propagator shows a rapid decrease in the region of $r \gsim 0.4$fm, which is related to the presence of the maximum in the dressing function and thus indicates the Gribov effects. On the other hand, the off-diagonal propagator shows less dependence for $\beta$. 

We have evaluated these behaviors qualitatively using the effective mass estimated from the linear slope of the logarithmic plot of $r^{1/2}G_{\mu\mu}(r)$ in the region of $r=0.4-0.8$fm, and compared it with the four dimensional result at long distance.
We have found that, in two dimensions,
the effective mass of off-diagonal gluons is estimated as $M_{\rm off} \simeq 1$ GeV, while   the mass of the diagonal gluon largely depends on $\beta$ and increases with increasing $\beta$. At $\beta=120$, the diagonal gluon mass becomes $M_{\rm diag} \simeq 2.2$ GeV, which is larger than the off-diagonal mass. On the other hand, in the four dimensions, the off-diagonal mass is estimated as $M_{\rm off} \simeq 1.6$ GeV, while the diagonal mass is as estimated as $M_{\rm diag} \simeq 0.3$ GeV. Therefore, this result does not only show that the Abelian dominance is not supported in the two dimensions and is supported in the four dimensions, but also indicates that the presence of monopole plays an important for the Abelian dominance. 

In addition, we have also calculated these propagators at zero-spatial-momentum, $D^0_{\mu \nu}(t)$. The longitudinal and transverse components correspond to $D^0_{4 4}(t)$ and $D^0_{11}(t)$, respectively. The logarithmic plots of all of these propagators show a curve which is concave downward and thus all of the spectral functions are expected to have negative regions, which means the violation of the Kallen-Lehmann representation. 

\section*{Acknowledgements}
The author thanks Daniel Zwanziger for useful discussions. 
This work is supported in part by a Grant-in-Aid for JSPS Fellows (No.24-1458)
from the Ministry of Education, Culture, Science and Technology 
(MEXT) of Japan.
The lattice QCD calculations are done on NEC SX-8R at Osaka University.

\appendix
\section{$d$-dimensional massive propagator in the Proca formalism}
\label{ap1}
In this appendix, we show the propagator based on the $d$-dimensional Lagrangian of 
the free massive vector field $A_\mu$ with the mass $M \ne 0$ 
in the Proca formalism,
\begin{align}
{\cal L}&= \frac{1}{4}(\partial_\mu A_\nu - \partial _\nu A_\mu)^2
	+\frac{1}{2}M^2A_\mu A_\mu,   \label{eqn:Lag} 
\end{align}
in the Euclidean metric. 
The propagator in momentum space $\tilde{G}_{\mu\nu}(p;M)$  is given by
\begin{align}
\tilde{G}_{\mu\nu}(p;M) = \left( \delta_{\mu\nu} -\frac{p_\mu p_\nu}{p^2}\right) \frac{1}{p^2+M^2} +\frac{1}{M^2} \frac{p_\mu p_\nu}{p^2}.
\end{align}
The longitudinal component $L(p^2)$ and the transverse component $T(p^2)$ defined in Sec.\ref{2} are regarded as
\begin{align}
 L(p^2) = \frac{1}{M^2},~ T(p^2) = \frac{1}{p^2 + M^2}. \label{LandT}
 \end{align}
Thus, the effective gluon mass is estimated by comparing the lattice QCD data of $L(p^2)$ and $T(p^2)$ with Eq. (\ref{LandT}).

Using this form, the scalar combination of the propagator in coordinate space $G_{\mu\mu}(r;M)$ reduces to 
\begin{align}
{G}_{\mu\mu}(r;M) &= \left< A_\mu(x) A_\mu(y) \right> \notag \\
	&=\int\frac{d^d k}{(2\pi)^d} e^{i k \cdot (x-y)}
	 \frac{1}{k^2+M^2}
 	\left( d+\frac{k^2}{M^2}\right) \notag\\
	&=
	\left(d-1\right)\int\frac{d^dk}{(2\pi)^d} e^{i k \cdot (x-y)}\frac{1}{k^2+M^2}    
	 + \frac{1}{M^2}{\delta^d(x-y)}\notag \\
	&=   \frac{d-1}{2\pi}\left(\frac{M}{2\pi r}\right)^{\frac{d}{2}-1}K_{d/2-1}(Mr)
+ \frac{1}{M^2}{\delta^d(x-y)}\notag \\
\label{eqn:prp02}
\end{align}
 with the modified Bessel function of the second kind, 
\begin{align}
K_\nu(z) = \frac{\sqrt{\pi}(z/2)^\nu}{\Gamma \left(\nu + \frac{1}{2}\right)}\int _1 ^\infty dt e^{-zt}\left(t^2-1 \right)^{\nu -1/2}.
\end{align}
In the infrared region with large $Mr$, 
Eq. (\ref{eqn:prp02}) reduces to 
\begin{align}
G_{\mu\mu}(r;M)
	&\simeq
      \frac{d-1}{2M}\left(\frac{M}{2\pi r}\right) ^{\frac{d-1}{2}}e^{-Mr}
\end{align}
using the asymptotic expansion, 
\begin{eqnarray}
K_\nu(z) \simeq
	\sqrt{\frac{\pi}{2z}} e^{-z} 
	\sum^\infty_{n=0}
	\frac{\Gamma(\frac{1}{2}+\nu+n)}{n!\Gamma(\frac{1}{2}+\nu-n)} \frac{1}{(2z)^n}, 
						\label{eqn:prp03A}
\end{eqnarray}
for large ${\rm Re}~z$ with $\Gamma$ the gamma function. Therefore  the effective gluon masses of the two dimensional gluon propagators are obtained by estimating the slope of the lattice QCD data of $\ln\left(r^{1/2}G_{\mu\mu}(r)\right)$ in coordinate space.

\bibliography{2DSU2MAL_Gprop}
\end{document}